\documentclass[preprint,aps]{revtex4}
\usepackage{graphicx}
\usepackage{mathptmx} 

\begin{document}






\noindent
{{\bf Comment on ``Do Earthquakes Exhibit Self-Organized Criticality?''}}

\vspace{0.3cm}

In a recent Letter, 
Yang, Du, and Ma \cite{Yang}
study the interesting problem 
of the temporal structure of seismicity and its
relation with self-organized criticality (SOC).
Their main finding is that the reshuffling 
of earthquake magnitudes
changes the shape of the earthquake recurrence-time 
(or first-return-time)
distribution when the low-magnitude bound, $M_c$,
is raised.
Subsequently, they conclude that 
it is not true that {\it an earthquake cannot ``know''
how large it will become}.
First, we show that this important implication is unjustified.

Yang {\it et al.} have in mind a fully uncorrelated
temporal point process with independent magnitudes as a picture
of SOC systems. 
It is obvious, by construction, 
that 
this model is invariant under
random rearrangements of the data; as
Yang {\it et al.} do not find this invariance in Southern California 
seisimicity they claim that  
``earthquakes do not happen with completely random magnitudes''
and therefore they are not a SOC phenomenon.
In fact, {\it the only conclusion that can be drawn from this is
that the seismicity time series is not uncorrelated}, and there
exists some dependence between magnitudes and recurrence times.
[This conclusion can be obtained directly, 
from the fact that a scaling law exists for the 
recurrence-time distributions corresponding to different 
low-magnitude bounds, with a scaling function that is not a decreasing
exponential \cite{Corral04} (characteristic of a Poisson process, the only 
uncorrelated process which verifies a scaling law).]

The existence of correlations 
means that,
for a given event $i$, its magnitude $M_i$ may depend on
the magnitude of the previous event, $M_{i-1}$,
as well as on the backwards recurrence time, $T_i=t_i-t_{i-1}$, 
with $t_i$ and $t_{i-1}$ the time of occurrence of both events.
This dependence can be extended to previous magnitudes and 
recurrence times, $T_{i-1}, M_{i-2}, T_{i-2},$ etc.
But further, the recurrence time to the next event, $T_{i+1}$,
may depend on the previous magnitudes, $M_j$ 
and recurrence times $T_j$, $j \le i$.
The reshuffling of magnitudes performed in Ref. \cite{Yang}
breaks (if they exist) the possible correlations of $M_i$ 
with the previous magnitudes, as well as with the previous 
recurrence times, and the correlations of $T_{i+1}$ with the 
previous magnitudes (but not with the previous recurrence times). 
Therefore, any of the influences $M_{i-1}\rightarrow M_i$,
$T_i\rightarrow M_i$, 
or $M_{i}\rightarrow T_{i+1}$,
may be responsible of Yang {\it et al.}'s results.

The most direct way to test the dependence of a given variable,
in this case $M_i$, with another variable $X$, 
is to measure
the probability density of $X$ conditioned to different values
of $M_i$, $P(X|M_i)$, and compare with the unconditioned probability density
of $X$, $P(X)$. 
This is what Fig. 1(a) displays, using $X=T_i$ and $X=T_{i+1}$
for the same data as Ref. \cite{Yang},
but restricted to periods of stationary seismicity
(otherwise, for strong aftershock sequences the recurrence times
are shorter and more sensitive to catalog incompleteness). 
As $P(T_i|M_i)$ remains practically unchanged for different sets of values of 
$M_i$, temporal causality leads to the conclusion that $M_i$ is independent
on $T_i$. 
In contrast, $T_{i+1}$ clearly depends on $M_i$, as $P(T_{i+1}|M_i)$ 
changes for different sets of values of $M_i$.
In other words, 
{\it the larger the magnitude $M_i$, the shorter 
the time to the next event $T_{i+1}$, 
but the value of this time has no influence on the magnitude
of the event, $M_{i+1}$}.
On the other hand, 
Fig. 1(b)
shows that $P(M_i|M_{i-1})$ turns out to be indistinguishable
from $P(M_i)$, ensuring the independence of $M_i$ and $M_{i-1}$, $\forall i$
if the $T_i$'s are restricted to be larger than 33 min
(shorter periods of time are not reliable, 
due to data incompleteness).
So, {\it when an earthquake starts, its magnitude is undetermined}
(at least from the information available at the catalogs),
whereas the time to the next event decreases when that magnitude 
turns out to be large.

A second, independent point to clarify
is the identification of SOC with the total absence of correlations.
It is true that the BTW sandpile model displays an exponential distribution
of recurrence times, but SOC is much more diverse than the BTW model.
For instance, the Bak-Sneppen model or the Oslo-ricepile model are 
two well recognized examples of SOC with totally different recurrence-time
distributions.
Finally, it is necessary to stress that the concept of SOC 
(as it happens with chaos) does not exclude the possibility
of prediction, as Ref. [17] of Yang {\it et al.} clearly showed.
So, nothing in Ref. \cite{Yang} is against the SOC picture
of earthquakes.


\noindent
\'Alvaro Corral

\noindent
Departament de F\'\i sica, Facultat de Ci\`encies, 
Universitat Aut\`onoma de Barcelona,
E-08193 Bellaterra, Spain

\today




\begin{figure}
\centering
\includegraphics[width=3.5in]{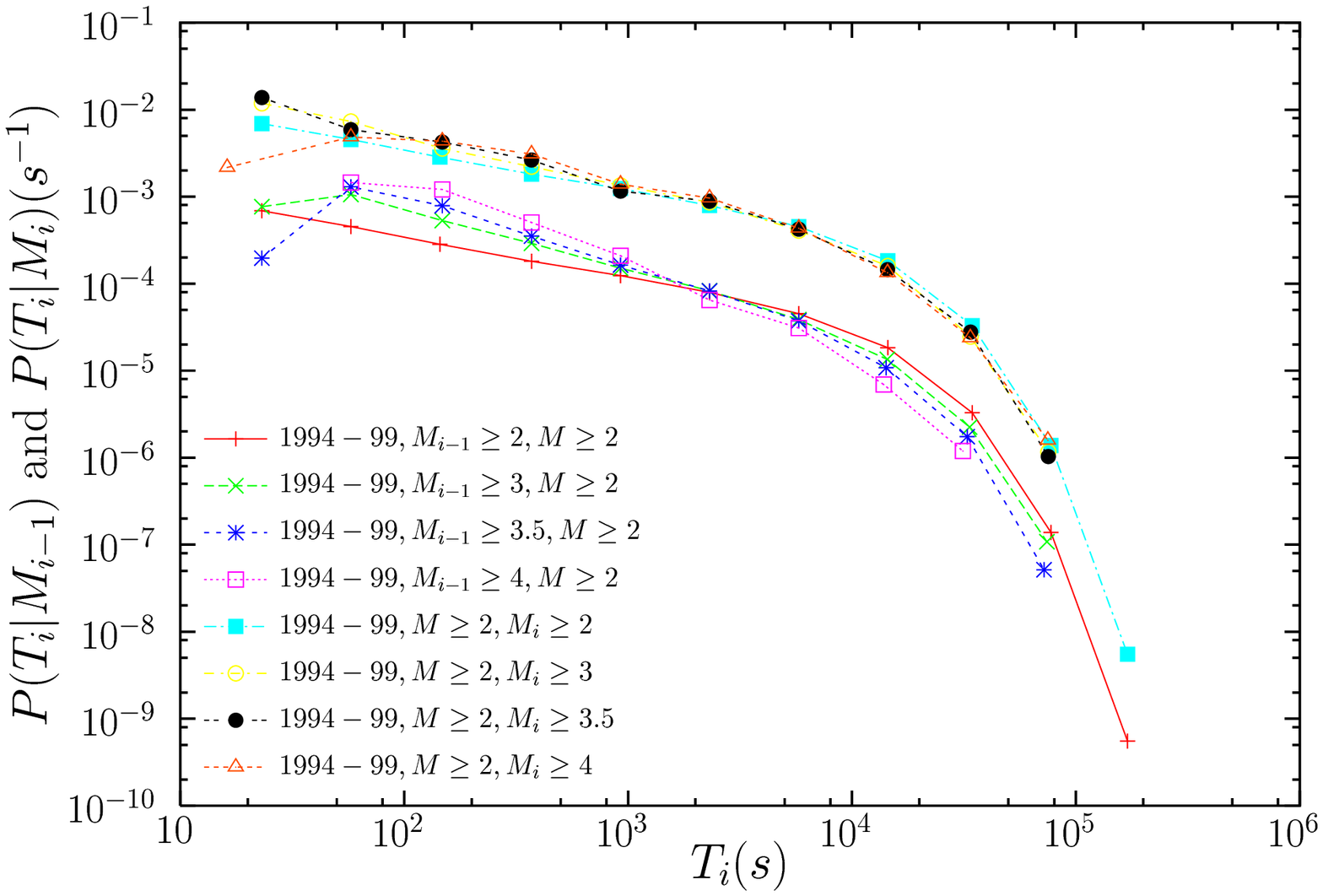}
\includegraphics[width=3.5in]{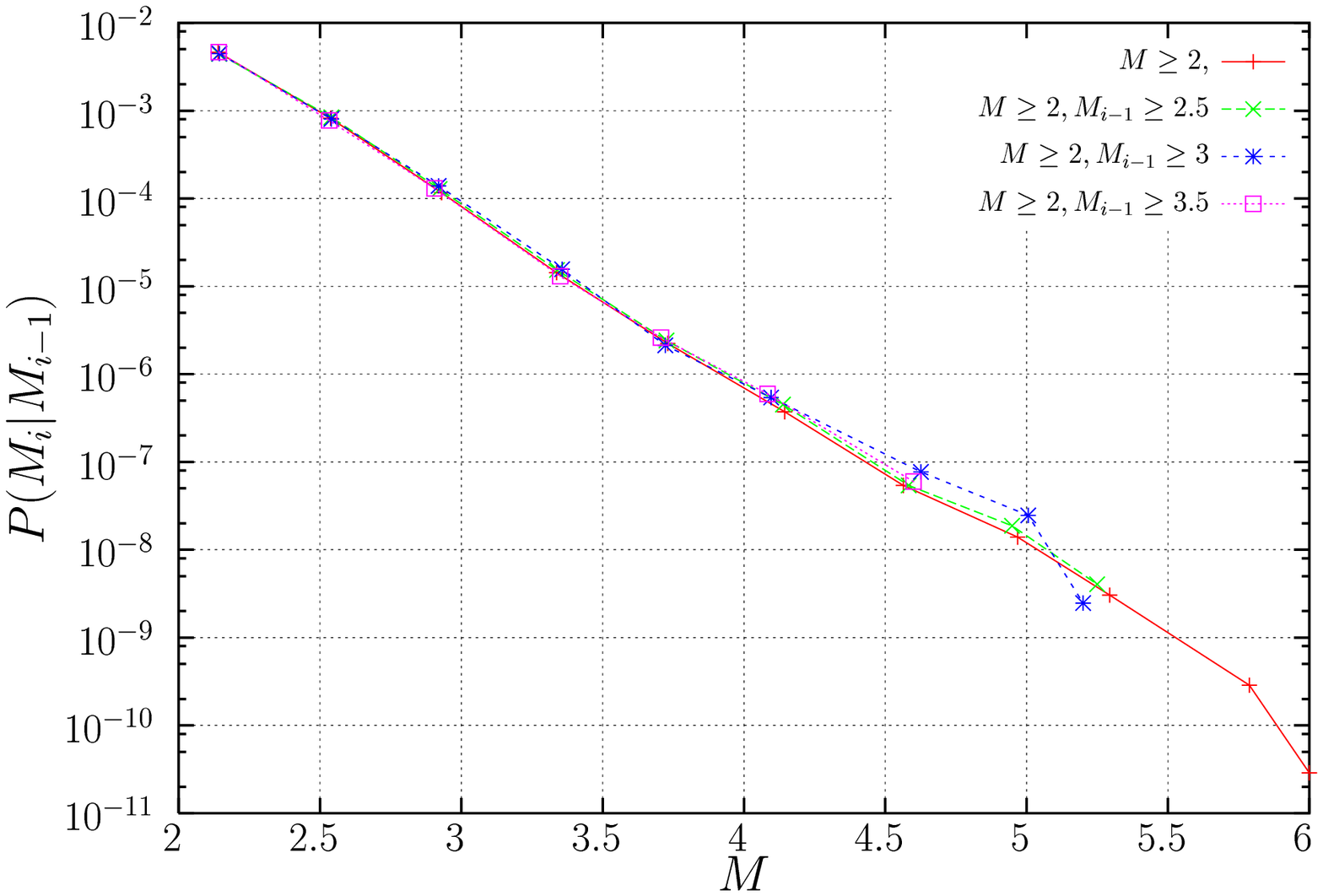}
\caption{
(color online)
(a) Probability densities $P(T_i|M_{i-1})$ and $P(T_i|M_{i})$
(shifted 
upwards) compared to $P(T_i)$.
(b) Probability density $P(M_i|M_{i-1})$
compared to $P(M_{i})$ with $T_i > 2000$ s.
\label{Dtcond}
}
\end{figure}

\end{document}